\documentclass[aps,prb,preprint,groupedaddress,showpacs]{revtex4}

\usepackage{graphics}

\begin{document}

\title{Wave Propagation Retrieval Method For Metamaterials: Unambiguous Restoration Of Effective Parameters}
\author{Andrei Andryieuski}
\email[]{andra@fotonik.dtu.dk}
\author{Radu Malureanu}
\author{Andrei V. Lavrinenko}
\affiliation{DTU Fotonik, Technical University of Denmark, Oersteds pl. 343, Kgs. Lyngby, DK-2800, Denmark}
\date{\today}

\begin{abstract}
In this article we propose a new direct method of effective parameters restoration that is based on the wave propagation phenomenon. It retrieves the effective properties unambiguously, is applicable to thick metamaterial (MTM) slabs and is easy in implementation. It is validated on the case studies of fishnet, split cube in carcass, Jerusalem cross and ultrahigh refractive index MTMs. The constraints of the method are designated.
\end{abstract}


\maketitle

\section{INTRODUCTION\label{Introduction}}
MTMs research is a new emerging field of photonics due to extravagant properties that can be obtained after relevant engineering of their structure. It provides great opportunities to control the light propagation. The potential applications of MTMs range from invisibility cloaks \cite{Schurig2005} to superresolution lenses \cite{Pendry2000} and nanocouplers \cite{Degiron2007}, but are not limited to them.

After some successful proof-of-principle experiments with the negative index behaviour, MTMs are moving now towards new frontiers, for example, having permittivity $\varepsilon$ or permeability $\mu$ either close to zero or very large, exhibiting isotropy or extreme anisotropy, dumping or enhancing spatial dispersion. In the cases when effective parameters (refractive index $n$, impedance $Z$, permittivity $\varepsilon$ and permeability $\mu$) can be introduced in the correct way, reliable retrieval procedure is on demand. Assigning some certain values for the parameters is the mean by which the properties of the MTM can be characterized in the universal mode.

The existing retrieval methods can be divided into following groups:

1) Reflection/transmission based. These methods are based on the reversal of the amplitude and phase reflection $r$ and transmission $t$ (or S-parameters) from the slab of homogenous material. Initially proposed for normal incidence on the isotropic metamaterial \cite{Smith2002, ChenX2004} the procedure was modified for inclined incidence \cite{Menzel2008A}, bianisotropic \cite{ChenX2005, Li2009} and chiral metamaterials \cite{Menzel2008B}. The procedure introduced by Smith et al. \cite{Smith2002} is easy to implement and it is widely used, so let we call it the standard method (SM).

SM suffers from the well-known "branch" problem. As we restore refractive index $Re(n)$ using the $\arccos$ function we have ambiguity of $2\pi m/(kd)$, where $m$ is an integer number, $k$ - wavenumber and $d$ - slab thickness. This method works well only for thin slabs. Unfortunately, a thin slab consisting from just a few MTM layers exhibits the properties which can be very different from the bulk MTM \cite{Rockstuhl2008}. Applying standard method to thick MTM slab is usually problematic. First, different branches of $Re(n)$ become very close to each other. Second, MTMs are usually absorptive so transmission through the thick slab is very low that leads through a numerical noise to large errors.

2) Field averaging \cite{Smith2006, Lerat2006}. In this case the fields $(E, D, H, B)$ in the unit cell are averaged on some surfaces and lines and the effective properties are found according to material relations $D= \varepsilon E$ and $B= \mu H$. Determining appropriate surfaces and contours for averaging is a non-trivial problem especially when optimizing 3D MTM structures.

3) Recently published quasimode theory \cite{Sun2009}. This method is based on the maximization of optical density of states for metamaterial while changing $\varepsilon$ and $\mu$ of the surrounding medium. The method is computationally demanding as it requires 4-parameters optimization for each frequency.

4) Based on the wave phenomenon. In this case the effective parameters are derived from observation of the wave propagation phenomenon. For example, Popa and Cummer \cite{Popa2005} proposed to match the simulated fields inside and outside of the thin metamaterial slab with the fields theoretically predicted by formulae. Fitting parameters in this case are refractive index and impedance. Due to multiple reflections the standing wave forms inside the slab so the fitting procedure is complicated.

The method we propose in this article is also based on the wave propagation phenomenon and we call it wave propagation retrieval method (WPRM). In some sense it exploits and elaborates ideas proposed in Ref. \cite{Popa2005} but WPRM has significant difference. We propose to restore the refractive index $n$ from the wave propagation phenomenon through very thick metamaterial slab. In this case we deal with only one wave propagating in the metamaterial so the retrieval formulae become easy to use. Such retrieval is unambiguous, and the effective parameters derived from the thick layer are bulk automatically. Impedance $z$ is restored from the input interface reflection only that leads to uniqueness of the retrieval and absence of effects caused by finiteness of the structure.

The article is organised as follow. In the part \ref{WPRMtheory} we give mathematical foundations of WPRM and systematically define the limits of its application. Case studies demonstrate the power of WPRM in the part \ref{CaseStudies}. Conclusions sum up the article.

\section{WAVE PROPAGATION RETRIEVAL METHOD\label{WPRMtheory}}
\subsection{Methodology\label{WPRMmethodology}}
\begin{figure}
 \includegraphics{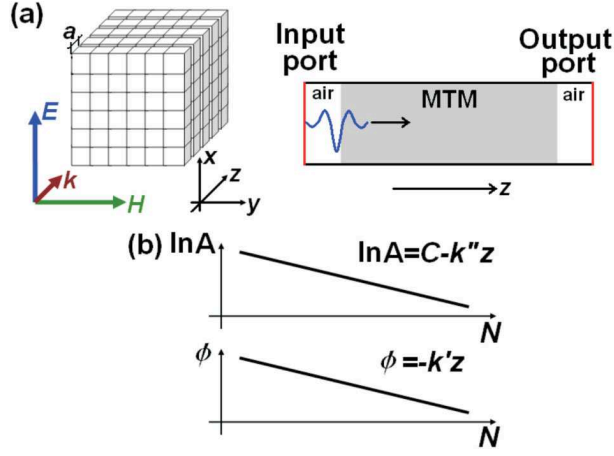}%
 \caption{Wave propagation simulation configuration (a); average amplitude and phase dependence on the unit cell number (theoretical expectation) (b).\label{FigSimulation}}
 \end{figure}

The simulation geometry assumes the MTM slab infinite in $x-$ and $y-$directions and semi-infinite in $z-$direction. The incidence of the wave occurs from vacuum to MTM. Inside MTM there is only one plane wave propagating along $z-$axis (Fig. \ref{FigSimulation}(a)). The fields have complicated spatial distribution inside the unit cell, so in order to apply an effective medium description we should abstract out from composite internal structure and regard it as a bulk homogeneous material. Thus we average the components of complex field vectors over the unit cell. Our choice is the electric field, but the procedure can be easily reformulated for the magnetic field. One more assumption we utilize here for the sake of simplicity is the linear polarization of the fields, which lead to the scalar procedure of restoration.

As we are working in the effective medium approach we describe the light wave of frequency $\omega$ as a modulated plane wave:
\begin{equation}
  E(x,y,z)=E_{0}u(x,y,z)exp(-ik'z-k''z),
 \label{equation1}
\end{equation}
where $u(x,y,z)$ is a modulation function periodic in $z-$direction with the same period $a$ as MTM, $k'=n'\omega/c$ and $k''=n''\omega/c$ are real and imaginary part of the wavenumber connected with the real $n'$ and imaginary $n"$ parts of the refractive index. Modulation function is normalized over the unit cell volume $V$: $\int u(x,y,z)dxdydz/V=1$. It changes quickly over the unit cell comparing with the exponential ansatz in (\ref{equation1}). Assuming the slow variation of the phase, which is valid under the generally accepted homogeneity conditions $\lambda_{eff}>4a$, the averaged over the $N-$th unit cell electric field is equal to
\begin{equation}
  \langle E\rangle_{N}=V^{-1}\int E_{0}u(x,y,z)exp(-ik'z-k''z)dxdydz \approx E_{0}exp(-k''z_N)exp(-ik'z_N)=Aexp(-i\phi),
 \label{equation2}
\end{equation}
where $z_{N}=(N-1/2)a$ is the coordinate of the $N-$th unit cell centre, $A=|\langle E\rangle_{N}|$ and $\phi=\arg(\langle E\rangle_{N})$ are the averaged field amplitude and phase, and the integration is performed over unit cell's volume. 

Both parts of the refractive index are straightforwardly derived via integral changes in phase $\Delta\phi$ and amplitude logarithm $\Delta\ln A$ over the fixed number of unit cells $\Delta N$:
\begin{equation}
  n'=-\frac{\Delta \phi}{\Delta N}\frac{c}{a\omega},
 \label{equation3}
\end{equation}
\begin{equation}
  n''=-\frac{\Delta \ln A}{\Delta N}\frac{c}{a\omega}.
 \label{equation4}
\end{equation}

Semi-infinite in $z-$direction configuration means that there is only one interface between vacuum and MTM. In this case impedance $Z$ can be restored in the unique way from the reflection $r$:
\begin{equation}
  Z=\frac{1+r}{1-r}.
 \label{equation5}
\end{equation}

It was mentioned by Landau \cite{Landau1984} that knowledge of the impedance for absorbing medium allows determining the fields outside it. In our case we solve the reverse problem determining impedance from the external fields. Knowing the refractive index and impedance one can easily obtain permittivity and permeability $\varepsilon=n/Z, \mu=nZ$.

\begin{figure}
 \includegraphics{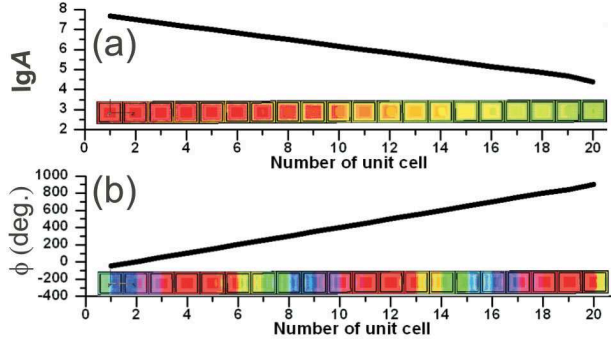}%
 \caption{An example of (a) amplitude and (b) phase of the averaged field distribution over 20 unit cells of the Split-Cube-in-Carcass design for the negative refractive index regime. Colour scales used in both cases for demonstration of the averaged amplitudes and phases are not equivalent.\label{WPRMexample}}
 \end{figure}

We performed the simulations of the wave propagation inside the thick MTM slab (20-50 MTM layers) with CST Microwave Studio 2009 (CST Computer Simulation Technology AG, Darmstadt, Germany) with a time-domain solver. For broadband excitation short modulated Gaussian pulse was chosen. As a reference we used the effective parameters retrieved with a SM from the reflection/transmission through 3 MTM layers simulated with a frequency-domain solver in CST. We found out that results retrieved with WPRM are the same as retrieved with SM for the pulse relative bandwidth up to 100\% (see part \ref{URIMTM}). Averaging and retrieval according to the formulae \ref{equation2}-\ref{equation5} were made with a homemade Matlab program. An example of WPRM realization for a split-cube in carcass (SCiC) structure (design description is below in the text) is shown on the Figure \ref{WPRMexample}.

\subsection{WPRM advantages and limitations\label{WPRMadvantages}}
WPRM has several clear advantages. First, it does not experience the "branch" problem. Second, WPRM does not suffer from low transmission and numerical noise as it uses the amplitude and phase distribution inside the structure - not only on the first and last slab interfaces. The method works well for thick slabs. Third, restoring the effective parameters with SM one should always use the broadband excitation source. There is no difference for WPRM whether to use broadband or narrowband source, so one can choose what excitation is more appropriate for a given problem.

\begin{figure}
 \includegraphics{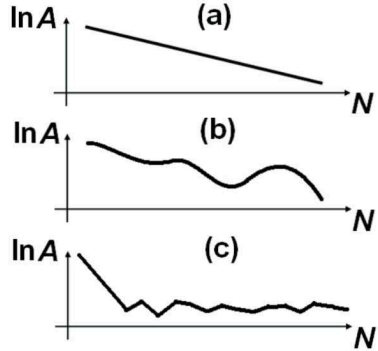}%
 \caption{\label{WPRMcases} Possible cases of simulated average amplitude distribution inside the MTM slab: (a) normal case; (b) standing wave formation due to the back interface reflection and slow wave attenuation; (c) large wave attenuation and numerical noise.}
 \end{figure}

WPRM also has some limitations. We cannot simulate semi-infinite MTM in practice so the reflection from the back interface may happen and lead to the standing wave formation. Different amplitude $\ln A$ distributions can occur except the most desirable linear one (Fig. \ref{WPRMcases}(a)) when formula (\ref{equation4}) works well. In the case of low wave attenuation and reflection from the back interface we can observe a standing wave pattern (Fig. \ref{WPRMcases}(b)). To retrieve $n$ and $Z$ correctly we should make the MTM slab thicker and take into account only the first linear part of the graph. In the case of large wave attenuation (Fig. \ref{WPRMcases}(c)) the amplitude value drops down within a short distance to the level of numerical noise. Noise data should be excluded for restoration.

\begin{figure}
 \includegraphics{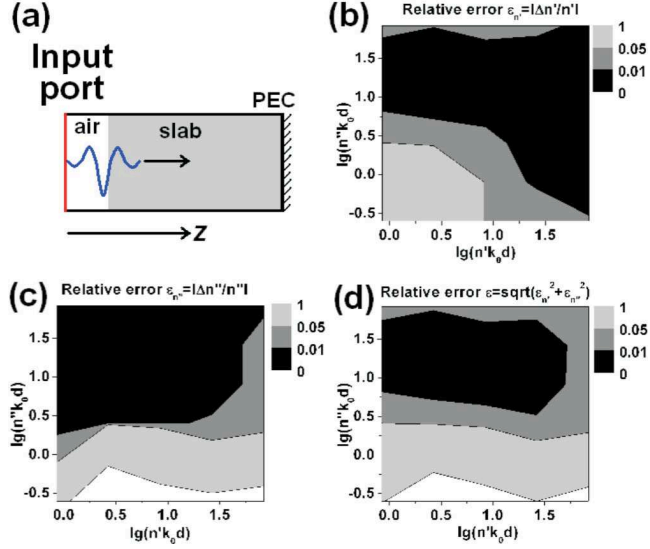}%
 \caption{\label{RelativeErrors} Configuration of "the worst case" simulation (a). Relative error of (b) $n'$, (c) $n"$ restoration and (d) total relative error. Black area correspond to the relative error less than 1\%. }
 \end{figure}

To find out the limitations of WPRM we simulated "the worst case": a homogenous slab with pre-defined refractive index $n$ and impedance $Z=1$ with a perfect mirror (PEC boundary condition) on the other side of the slab (Fig. \ref{RelativeErrors}(a)). Numerical simulations were performed in CST until the total electromagnetic energy stored in the system dropped down to -80dB with respect to the maximum level.

It is obvious that not the value of the refractive index itself but the terms in the exponentials in (\ref{equation2}) play the role in phase and amplitude changes. So we calculated relative error of $n$ restoration for different $n'k_{0}d$ and $n"k_{0}d$ (Fig. \ref{RelativeErrors}(b)-(d)), where $k_0=\omega/c$ is the wavenumber in vacuum. In the case of fast amplitude decay (Fig. \ref{WPRMcases}(c)) we should consider $d$ not the full thickness of the slab but only until the $\ln A$ reaches the noise level. Thou different regions correspond to the small restoration error in $n'$ and $n''$ (Fig. \ref{RelativeErrors} (b) and (c)), there is an optimal region where the total relative error is less than 1\% (Fig. \ref{RelativeErrors}(d)), which can be approximated with a rectangle. We checked WPRM for positive index materials but the results are applicable for the negative index materials as well. So we might state that WPRM is applicable with a relative error less than 1\% for $-45\leq n'k_{0}d\leq 45$ and $5\leq n"k_{0}d\leq 56$. In the usual MTM cases these limits are wider as we do not have perfect reflector at the second interface.

The range of applicability of WPRM can be extended to the case of low-loss structures if we excite the MTM slab with short Gaussian pulse and stop the simulation when the centre of the pulse reaches the second interface of the slab. In this case we have a reflected wave only near the second interface and can skip these amplitude and phase data. As a shortcoming such simulation is possible only for metamaterials with low group velocity dispersion and requires additional estimation of the time the pulse travels through the slab. The case studies of applicability of WPRM to low-loss MTMs follow below in the sections \ref{JerusalemCrosses} and \ref{URIMTM}.

\section{CASE STUDIES\label{CaseStudies}}
To check how good WPRM works we choose MTM designs for different electromagnetic range: optical, terahertz and microwave. Standard method \cite{Smith2002} was used as a reference.
\subsection{Fishnet\label{Fishnet}}
\begin{figure}
 \includegraphics{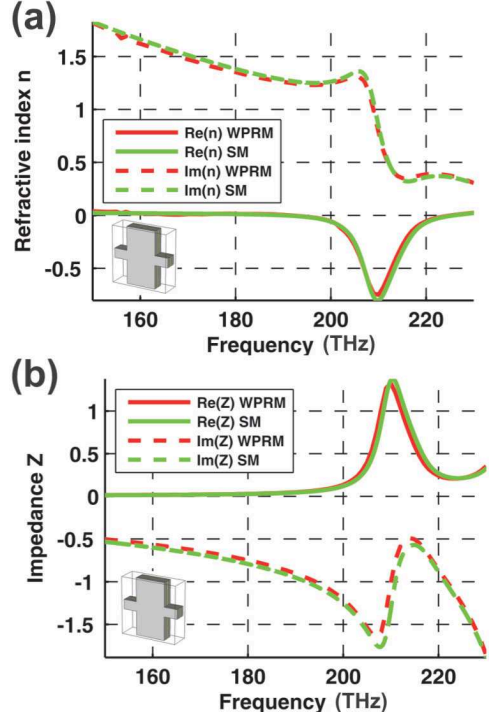}
 \caption{\label{FigFishnet}Refractive index (a) and impedance (b) of the fishnet structure restored with WPRM (red line) and SM (green line). Real and imaginary parts are marked with solid and dashed lines respectively. This colour and symbol convention will be respected in all subsequent figures.}
 \end{figure}

Fishnet structure \cite{Dolling2006} is a classical example of the optical negative index MTM with high transmittivity. For WPRM test we took a specific design parameters from \cite{ Rockstuhl2008}. It was shown in \cite{Rockstuhl2008} the dominance of a single Bloch mode and fast convergence of the effective parameters with a slab thickness.
Effective parameters $n$ and $Z$ restored with WPRM and SM are in a perfect agreement (Fig. \ref{FigFishnet}).

\subsection{Split cube in carcass\label{SCiC}}
The split cube in carcass structure (SCiC) is similar to the split cube in cage design \cite{Andryieuski2009}. Its remarkable property is bulkiness: the effective parameters at negative index region do not depend on the number of layers starting from a single layer \cite{Andryieuski2009}. The structure possesses cubic symmetry that insures its polarization insensitivity. As a drawback, SCiC has quite low transmittivity and high absorption.

\begin{figure}
 \includegraphics{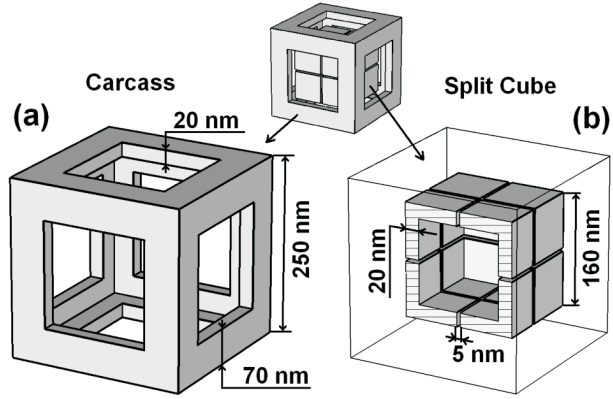}%
 \caption{\label{SCiCdesign}Split Cube in Carcass unit cell design: (a) Carcass, (b) Split Cube.}
 \end{figure}

The SCiC unit cell consists of two silver parts embedded in silica (refractive index $n=1.5$). Silver is regarded as Drude metal with plasma frequency $\omega_p=1.37\times10^{16}$ rad/s and collision frequency $\nu_c=8.50\times10^{13}s^{-1}$ \cite{Dolling2006}. The outer part carcass (Fig. \ref{SCiCdesign} (a)) is a sort of 3D wire medium and provides negative $\varepsilon'$. The inner part split cube (Fig. \ref{SCiCdesign} (b)) is the hollow cube with the slits in the middle of the facets. It is the logical 3D extension of the split ring resonator concept and gives negative $\mu'$. The details about structure's sizes are labeled on the Figure \ref{SCiCdesign}.

\begin{figure}
 \includegraphics{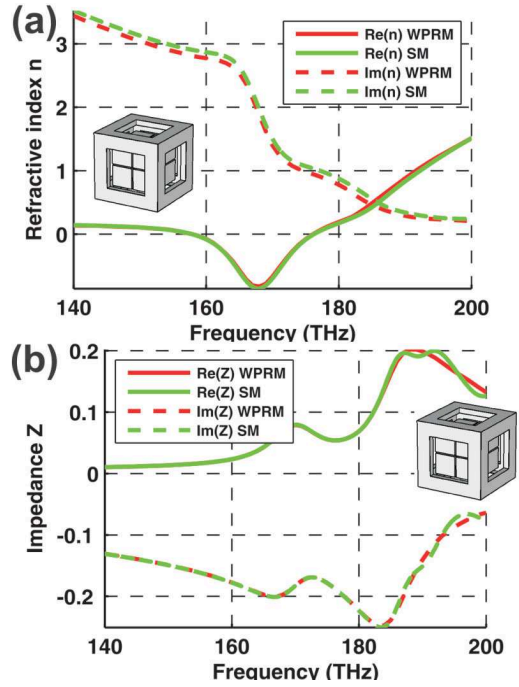}%
 \caption{\label{FigSCiC}Split Cube in Carcass effective parameters: (a) refractive index and (b) impedance.}
 \end{figure}

Both WPRM and SM give very similar results (Fig. \ref{FigSCiC}). For the frequencies above 190THz there are differences in the retrieved impedance (Fig. \ref{FigSCiC} (b)). They come out from the fact of the dipole resonances of split cubes and failure of the effective homogeneity approach \cite{Andryieuski2009}.

\subsection{Jerusalem crosses\label{JerusalemCrosses}}
Recently Wang et al. reported a Jerusalem crosses 3D MTM design for microwave region claiming isotropic broadband left-handedness \cite{Wang2009}. As the reported losses in this MTM are relatively low it was a good opportunity to check WPRM. All the design parameters were taken the same as in \cite{Wang2009}.

\begin{figure}
 \includegraphics{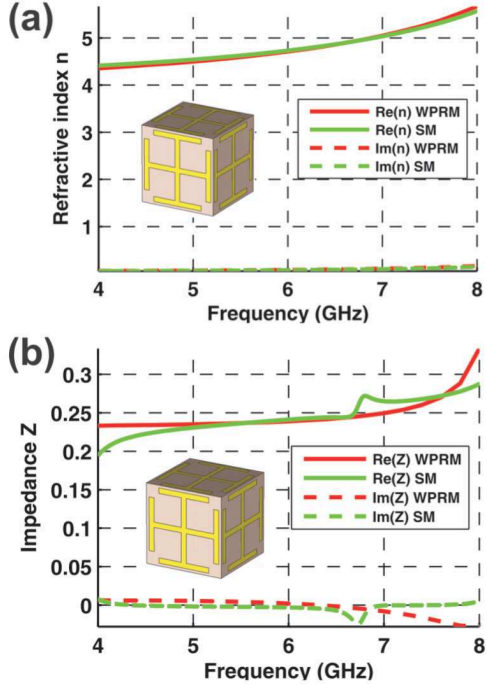}%
 \caption{\label{FigJerusalemCrosses}Jerusalem crosses MTM effective parameters: (a) refractive index and (b) impedance.}
 \end{figure}

To our astonishment WPRM showed no negative refractive index in the specified region (around 6GHz) (Fig. \ref{FigJerusalemCrosses}(a)). WPRM results perfectly coincide with the SM results if one chooses the correct branch. These results are feasible as real and imaginary parts of $n$ must obey the Kramers-Kronig relations \cite{Lucarini2005}. Small imaginary part corresponds to smooth changes in the real part in contrary to \cite{Wang2009}. This is another confirmation that WPRM allows to avoid errors and misunderstanding in the refractive index restoration and interpretation.

A kink in the impedance spectrum restored with SM (Fig. \ref{FigJerusalemCrosses}(b)) appears due to the finiteness of the system. Additional studies showed that it changes its position with the number of layers always being exactly at the frequency of Fabry-Perot resonance. There is no such problem for WPRM as multiple reflections from the slab interfaces are successfully avoided.

\subsection{3D metamaterial with ultrahigh refractive index\label{URIMTM}}
\begin{figure}
 \includegraphics{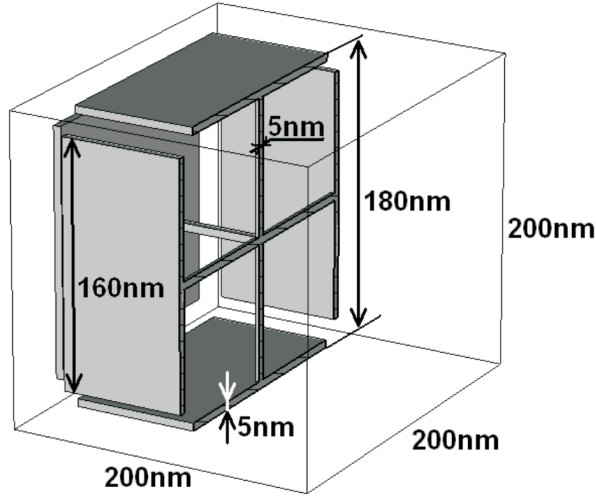}%
 \caption{\label{URIMTMdesign} URIMTM design geometrical parameters. Only half of the unit cell is shown. Metallic structure is embedded in silicon.}
 \end{figure}

A 3D MTM with ultrahigh refractive index (URIMTM) over a broad bandwidth has been recently proposed by Shin et al. \cite{Shin2009}. Their idea is to use unit cell consisting of metallic plates connected with wires to obtain the refractive index up to 7. In \cite{Shin2009} there were not specified the exact geometrical parameters of the unit cell so we constructed a similar structure (Fig.~\ref{URIMTMdesign}) from silver (metal parameters the same as in part \ref{SCiC}) embedded in silicon ($\varepsilon=12$).

 \begin{figure}
 \includegraphics{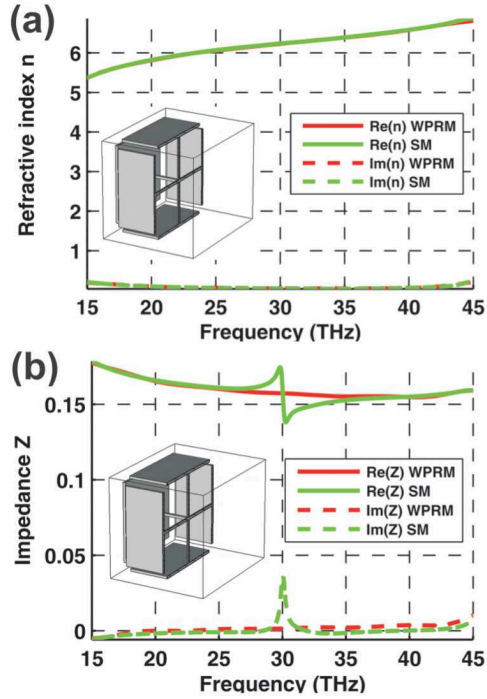}%
 \caption{\label{FigURIMTM} URIMTM effective parameters: (a) refractive index and (b) impedance.}
 \end{figure}

URIMTM shows high refractive index up to 7 in a broad range of frequencies (Fig.~\ref{FigURIMTM}(a)). WPRM and SM results are very close to each other. A kink in the impedance around 30 THz (Fig.~\ref{FigURIMTM}(b)) is due to finiteness of the structure as it is discusses in the part \ref{JerusalemCrosses}.

It is worth to mention that $\lg(n"k_{0}d)=-0.75$ at frequency 30THz, which is out of the low error limits defined in Figure \ref{RelativeErrors}(d), but relative difference of WPRM with SM is only 7\%. This fact confirms applicability of WPRM to low-loss MTM as well as to low-loss.

\section{CONCLUSIONS\label{Conclusions}}
In this work we proposed a new effective parameters retrieval method based on the wave propagation phenomenon and validate its applicability to lossy and low-loss metamaterials on the case studies. The main advantage of WPRM is that it retrieves the effective properties unambiguously and is applicable to a thick MTM slab. WPRM does not depend on the working frequency so it can be used in microwaves, terahertz and optical regions - while an effective medium approximation is valid. WPRM is easy to implement using commercially available electromagnetic simulation software.

In principle, WPRM can be extended to anisotropic media or media with a non-local response as soon as required observable features of the wave can be detected numerically. 

\begin{acknowledgments}
The authors gratefully acknowledge partial supports from the Danish Research Council for Technology and Production Sciences via the NIMbus project and from COST Action MP0702.
\end{acknowledgments}

\bibliography{Andryieuski}

\end{document}